\documentclass[secnumarabic, graphics,floatfix, nofootinbib,tightenlines,nobibnotes, aps, prl, 12pt]{revtex4-1}
\usepackage{graphicx}
\usepackage[english]{babel}
\usepackage[utf8]{inputenc}
\usepackage{amsmath,amssymb}
\usepackage{amsfonts}
\usepackage{multirow}
\usepackage{pstricks}
\usepackage{float}
\usepackage{subfigure}
\usepackage{color}
\usepackage{epsfig}
\usepackage{pst-tools}

\begin{document}

\title{Gravitational Waves in Scalar-Tensor-Vector Gravity Theory}
\author{Yunqi Liu$^{1,2} $, Wei-Liang Qian$^{2,3,1}$, Yungui Gong$^{4}$, and Bin Wang$^{1,5}$}
\affiliation{
$^{1}$ Center for Gravitation and Cosmology, College of Physical Science and Technology,Yangzhou University, Yangzhou 225009, China\\
$^{2}$ Escola de Engenharia de Lorena, Universidade de S\~ao Paulo, 12602-810, Lorena, SP, Brazil\\
$^{3}$ Faculdade de Engenharia de Guaratinguet\'a, Universidade Estadual Paulista, 12516-410, Guaratinguet\'a, SP, Brazil\\
$^{4}$ School of Physics, Huazhong University of Science and Technology, 430074, Wuhan, Hubei, China\\
$^{5}$ Collaborative Innovation Center of IFSA (CICIFSA), Shanghai Jiao Tong University, 200240, Shanghai, China\\
}

%\date{\today}
\date{Dec. 17, 2019}

\begin{abstract}
In this paper, we study the properties of gravitational waves in the scalar-tensor-vector gravity theory.
The polarizations of the gravitational waves are investigated by analyzing the relative motion of the test particles. 
It is found that the interaction between the matter and vector field in the theory leads to two additional transverse polarization modes. 
By making use of the polarization content, the stress-energy pseudo-tensor is calculated by employing the perturbed equation method. 
Besides, the relaxed field equation for the modified gravity in question is derived by using the Landau-Lifshitz formalism suitable to systems with non-negligible self-gravity.

\end{abstract}

\maketitle

\section{I. Introduction}\label{introduction}

The recent observation of gravitational waves (GWs) by the LIGO and Virgo Scientific Collaboration opens up a new avenue to explore the gravitational physics from an entirely new perspective~\cite{LIGO-Virgo1,LIGO-Virgo2,LIGO-Virgo3,LIGO-Virgo4,LIGO-Virgo5,LIGO-Virgo6,LIGO-Virgo7}. 
As a matter of fact, the direct detection of gravitational waves is an extremely elaborated process due to the smallness of the effect that the waves produce. 
In the case of the resonant mass antennas and interferometers, that most GW detectors employ, precise measurements critically rely on the rate of change regarding the GW phase and frequency. 
The latter, in turn, may depend on the specific theory of gravity.
To be specific, a particular choice of the theory of gravity entails subtlety and how the relevant information is separated from the noise.
Subsequently, it affects how the physical content is interpreted.

Moreover, comparing to the plus and cross polarizations in Einstein's General Relativity, alternative metric theories of gravity usually possess more polarizations associated with additional degrees of freedom~\cite{Eardley1973,Capozziello2013,will2014,Liang-Gong-Hou-Liu,hou-gong-liu,gong-hou,jacobson-mattingly,seifert}. 
Therefore, the measurement of distinct polarization modes of GWs can serve to discriminate between different gravitational theories~\cite{chatziioannou-yunes-cornish,LIGO-Virgo9,LIGO-Virgo10}.
If a vector or scalar polarization component is observed in GWs, it could provide strong evidence for modified gravity, as General Relativity only predicts tensorial polarizations.
In particular, the event GW170814 was, for the first time, utilized to extract the polarization character of the GWs~\cite{LIGO-Virgo4}.
The results are consistent with Einstein's gravity as the analysis indicated that the pure tensor polarizations are favored over the scenario of pure vector or scalar polarizations.
Nonetheless, the properties of the modified gravity regarding GWs have recently become a topic of increasing interest~\cite{gong2018,gong2019,zhao2017,zhao2018,zhao2019}.
On the theoretical side, several methods can be employed to analyze the polarization modes of gravity theory. 
Among others, one commonly employed approach is to study the polarization of weak, plane, and null GWs by using Newman-Penrose (NP) formalism~\cite{NP,Eardley-Lightman-Wagoner-Will,Nishizawa-Taruya-Kawamura-Sakagami,Alves-Miranda-Araujo,Myung-Moon,hou-gong-liu,Wagle-Saffer-Yunes}. 
Another method consists of reducing the metric into irreducible components and rewriting the linearized gravitational equations into the independent scalar, vector, and tensorial parts~\cite{Flanagan-Hughes,Poisson-Will}. 
Subsequently, the latter can be classified into radiative and non-radiative degrees of freedom. 
A third method is to investigate the relative motion of neighboring particles regarding the geodesic deviation equation in General Relativity~\cite{misner-thorne-wheeler}. 
In alternative theories, however, test particles do not necessarily move along geodesics.
Moreover, the relative acceleration of neighboring particles may also deviate from what is expected from General Relativity. 
Subsequently, the information on polarizations can be understood by placing test particles on a sphere around the observer, and studying how the sphere deforms in time~\cite{Swaminarayan-Safko83,hou-gong,hou-gong-liu}. 
This work will employ the third method to explore the polarization content.

The energy and momentum propagation is also an intriguing aspect in the study of GWs.
From the experimental viewpoint, the extraction of GW from the noise depends on accurate modeling of the rate of change of the energy of a gravitational system.
Theoretically, according to the balance law, the latter is identical to what emitted from a gravitational system via the GWs in terms of all possible degrees of freedom. 
The latter can be obtained via the GW stress-energy pseudo tensor (SET)~\cite{isaacson1,isaacson2,misner-thorne-wheeler}. 
With the information of polarizations for GWs at hand, it is possible to find the rate of energy by enumerating all propagating degrees of freedom. 
In literature, a variety of approaches have been developed and discussed~\cite{Nother,landau-lifshitz,stein-yunes}.
For a review, see Ref.~\cite{saffer-yunes-yagi}. 
The traditional perturbed field equation method consists of obtaining the equation of motion of small deviations of the metric from a generic background. 
In particular, the equation of motion of the second-order metric perturbations furnishes the GW SET by carrying out the short-wavelength average~\cite{isaacson1,isaacson2}.

In practice, the linearized theory is typically adopted to handle the generation of GW when the source's self-gravity has a negligible influence on its motions.
However, particular attention is required when the above condition does not hold. 
To be specific, for systems whose dynamics are dominated by self-gravity, even for the case of weak gravity, the linearized theory is no longer applicable. 
This was first pointed out by Eddington. 
Typical examples of such scenarios are binary star systems or those with nonlinear GW memory effect~\cite{payne,blanchet-damour,christodoulou,thorne}. 
In this case, one needs to elaborate on a specific approach for the system of the weak field but with non-negligible self-gravity~\cite{thorne1975,Flanagan-Hughes}. 
In General Relativity, such a formalism of the Einstein equations is known as ``relaxed Einstein equations''.
Therefore, in the present study, the resultant equation would be referred to as ``relaxed gravitational equation''. 
In Ref.~\cite{Nutku}, the author developed such a formalism for the Brans-Dicke theory.

Scalar-Tensor-Vector gravity (STVG) is an alternative metric theory that is characterized by the exchange of spin-zero and spin-one bosons. 
As a modified gravitational (MOG) theory~\cite{moffat2006}, it has been successfully applied to many contexts. 
The latter include solar system observations~\cite{moffat1410}, the rotation curves of galaxies~\cite{moffat2013,moffat043004}, the dynamics of galactic clusters~\cite{moffat-rahvar,Brownstein-Moffat}, description of the growth of structure, the matter power spectrum, as well as the cosmic microwave background (CMB) acoustical power spectrum data~\cite{Moffat1409}. 
In Ref.~\cite{moffat2013,moffat2014}, Moffat {\it et al.} investigated the weak field approximation and the constraints associated with the observed galaxy rotation curves and Chandra X-ray Clusters. 
In Ref.~\cite{green-moffat-toth}, the authors pointed out the existence of three gravitons in the STVG theory, which propagate at the speed of light and are in consistence with observations of the events GW170817/GRB170817A.
Therefore, it is intriguing to extend further the study of the STVG theory, which have, by and large, explored regarding week field applications, to the context of GW. 
Besides the properties of the GW, we are also interested in the ``relaxed gravitational equation'', which measures the deviation in the motion of neighboring particles from General Relativity.

The present study involves an attempt to investigate various aspects of the GWs in STVG gravity.
We explore the polarization modes of weak, plane GWs in STVG theory by the relative motion of test particles in Fermi normal coordinates. 
Also, the calculations of the SET are carried out by employing both the perturbed equation method and Landau-Lifshitz formalism. 
Moreover, the relaxed gravitational equations are derived for the system where self-gravity is not negligible.
The rest of the paper is organized as follows.  
In Section II, we present the formalism of the STVG.
The linearized field equations regarding small perturbations in the STVG theory is derived in Section III.
Section IV is dedicated to discussing the polarization of the perturbations by investigating the relative motion of neighboring particles in Fermi normal coordinates.
The calculations of the SET is carried out in Sections V and VI.
In Section V, the traditional perturbed field equations method is employed to obtain the SET.
The latter is compared against that obtained by utilizing the Landau-Lifshitz formulation in Section VI.
Also, the relaxed field equation is obtained and discussed. 
Concluding remarks are given in the last section. 
Throughout this work, we make use of the metric signature $(-,+,+,+)$.

\section{II. The STVG theory}\label{empt}
The generic form of the STVG action is given by~\cite{moffat2006, moffat2013, moffat2014}:
\begin{eqnarray}\label{action}
S=S_G+S_{\phi}+S_S+S_M.
\end{eqnarray}
It is composed of degrees of freedom of the metric $g_{\mu\nu}$ with the cosmological constant $\Lambda$, a vector field $\phi^{\mu}$ and dynamical scalar fields.
The latter consists of the gravitational coupling strength $G$ and the mass of the vector field $\mu$.
To be specific,
\begin{eqnarray}\label{actionterms}
S_G&=&\frac{1}{16\pi}\int d^4 x \sqrt{-g}\frac{1}{G}(R+2\Lambda),\\
S_{\phi}&=&\int d^4 x \sqrt{-g}\left[-\frac{1}{4}B^{\mu\nu}B_{\mu\nu}+\frac12 \mu^2\phi_\mu\phi^\mu+V_{\phi}\right],\\
S_S&=&\int d^4x\sqrt{-g}\left[\frac{1}{G^3}\left(\frac{1}{2}g^{\mu\nu}\partial_{\mu}G\partial_{\nu}G-V_G\right)+\frac{1}{\mu^2G}\left(\frac{1}{2}g^{\mu\nu}\partial_{\mu}\mu\partial_{\nu}\mu-V_{\mu}\right)\right] ,
\end{eqnarray}
where $B_{\mu\nu}=\partial_{\mu}\phi_{\nu}-\partial_{\nu}\phi_{\mu}$.
The self-potentials $V_{\phi}$, $V_G$, and $V_{\mu}$ are associated with the vector field and the scalar fields.

The field equations are given by,
\begin{eqnarray}\label{fieldsequations}
&&X_{\mu\nu}=G_{\mu\nu}-\Lambda g_{\mu\nu}+Q_{\mu\nu}-8\pi GT_{\mu\nu}=8\pi GT^{M}_{\mu\nu},\label{eqg}\\
&&\nabla _{\nu}B^{\mu\nu}+\frac{\partial V_{\phi}}{\partial \phi^{\mu}}=0,\label{eqv}\\ 
&&\square G=K,\label{eqG}\\
&&\square \mu=L,\label{eqmu}
\end{eqnarray}
where
\begin{eqnarray}\label{}
T_{\mu\nu}&=&T^{\phi}_{\mu\nu}+T^{G}_{\mu\nu}+T^{\mu}_{\mu\nu},\\
T^{\phi}_{\mu\nu}&=&-\frac{1}{4\pi}\left[B_{\mu}^{\alpha}B_{\nu\alpha}-g_{\mu\nu}(\frac{1}{4}B^{\rho\alpha}B_{\rho\alpha}+V_{\phi})\right],\\
T^{G}_{\mu\nu}&=&-\frac{1}{4\pi G^3}(G_{;\mu}G^{;\nu}-\frac{1}{2}g_{\mu\nu}G_{;\alpha}G^{;\alpha}),\\
T^{\mu}_{\mu\nu}&=&-\frac{1}{4\pi G \mu^2}(\mu_{;\mu}\mu_{;\nu}-\frac{1}{2}g_{\mu\nu}\mu_{;\alpha}\mu^{;\alpha}),\\
Q_{\mu\nu}&=&\frac{2}{G^2}(G^{;\alpha}G_{;\alpha}g_{\mu\nu}-G_{;\mu}G_{;\nu})-\frac{1}{G}(G_{;\alpha}^{;\alpha}g_{\mu\nu}-G_{;\mu;\nu}),\\
K&=&\frac{3}{G}(\frac{1}{2}G_{;\mu}G^{;\mu}+V_G)-\frac{G}{\mu^2}(\frac{1}{2}\mu_{;\mu}\mu^{;\mu}-V_{\mu})-\frac{\partial V_G}{\partial G}-\frac{G}{16\pi}R,\\
L&=&\frac{1}{G}G^{;\mu}\mu_{;\mu}+\frac{\mu_{;\mu}G^{;\mu}}{G}-G \mu^3\phi_{\mu}\phi^{\mu}+\frac{2}{\mu}V_{\mu}-\frac{\partial V_{\mu}}{\partial \mu}.
\end{eqnarray}
Here the Einstein tensor $G_{\mu\nu}=R_{\mu\nu}-\frac{1}{2}R g_{\mu\nu}$.
$T^{\phi}_{\mu\nu}$, $T^{G}_{\mu\nu}$, and $T^{\mu}_{\mu\nu}$ are the energy-momentum tensors for the fields $\phi^{\mu}$, $G$, and $\mu$, respectively. 
$T^{M}_{\mu\nu}$ represents the energy-momentum tensor of matter fields, the semicolon ``;'' denotes the covariant derivative compatible with the metric $g_{\mu\nu}$.
In the present study, both the cosmological constant $\Lambda$ and the energy-momentum tensor $T^{M}_{\mu\nu}$ are taken to be zero as we are only interested in the GWs in the flat vacuum spacetime.
For simplicity, one also assumes that self-interaction potentials $V_{\phi}$, $V_{G}$, and $V_{\mu}$ vanish in the following discussions.

\section{III. Linearized field equations regarding perturbations} \label{weak-field}

In this section, we derive the equations of motion of small perturbations under the weak field approximation.  
Following the strategy presented in Refs.~\cite{moffat2013,moffat2014}, one perturbs the metric around the Minkowski spacetime $\eta_{\mu\nu}$ to the first order
\begin{eqnarray}\label{metricpert}
g_{\mu\nu}=\eta_{\mu\nu}+h^{(1)}_{\mu\nu},
\end{eqnarray}
and similarly expands the scalar and vector fields, namely
\begin{eqnarray}\label{fieldpert}
G&=G^{(0)}+G^{(1)},\\
\mu&=\mu^{(0)}+\mu^{(1)},\\
\phi^{\mu}&=\phi_{\mu}^{(0)}+\phi_{\mu}^{(1)}.
\end{eqnarray}
In the above equations, the superscript ``$^{(0)}$'' indicates a zeroth order contribution, while ``$^{(1)}$'' denotes the first-order perturbations. 
Regarding the flat spacetime background, $G^{(0)}$ and $\mu^{(0)}$ are constants in time and $\phi_{\mu}^{(0)}$ vanishes. 
As a standard practice, linearized tensors are raised or lowered by the Minkowski metric $\eta_{\mu\nu}$.

Then, one linearizes Eq.~\eqref{eqg} and introduces the notation 
\begin{eqnarray}\label{newpert}
\overline{h}_{\mu\nu}=h^{(1)}_{\mu\nu}-\frac{1}{2}h^{(1)}\eta_{\mu\nu}+\psi\eta_{\mu\nu} ,
\end{eqnarray}
where one has rewritten the scalar perturbation as $\psi=G^{(1)}/G^{(0)}$. 
It is straightforward to see that Eq.~\eqref{newpert} implies 
\begin{eqnarray}\label{newpert-re}
h_{\mu\nu}^{(1)}=\overline{h}_{\mu\nu}-\frac{1}{2}\overline{h}\eta_{\mu\nu}+\psi\eta_{\mu\nu},
\end{eqnarray}
where $h^{(1)}$ and $\overline{h}$ are the traces of $h^{(1)}_{\mu\nu}$ and $\overline{h}_{\mu\nu}$, respectively.

Subsequently, by making use of Eq.~\eqref{newpert}, the equation of motion for $h^{(1)}_{\mu\nu}$ can be rewritten as
\begin{eqnarray}\label{gw0}
\overline{h}_{\alpha\beta,\gamma}^{~~~~~,\gamma}+\eta_{\alpha\beta}\overline{h}_{\gamma\delta}^{~~,\gamma\delta}-\overline{h}_{\beta\gamma}^{~~,\alpha\gamma}+\overline{h}_{\alpha\gamma}^{~~,\beta\gamma}=0.
\end{eqnarray}
where the comma ``,'' denotes the ordinary derivative with respect to the metric $\eta_{\mu\nu}$.
We observe that Eq.~\eqref{gw0} possesses the same form as in General Relativity.
Therefore, one can similarly impose the Lorenz gauge $\partial^{\mu}\overline{h}_{\mu\nu}=0$ , and Eq.~\eqref{gw0} is thus simplified to 
\begin{eqnarray}\label{gw}
\overline{h}_{\alpha\beta,\gamma}^{~~~~~,\gamma}=0.
\end{eqnarray}
It is also noted that the Lorenz condition does not entirely fix the gauge freedom, one can utilize the residual symmetry to require $\overline{h}=0$. This is called the transverse-traceless gauge, or TT gauge. 

The plane wave solution of Eq.~\eqref{gw} can be expressed as 
\begin{eqnarray}\label{solutiontogw}
\overline{h}_{\alpha\beta}=C_{\alpha\beta} \exp(i q_{\mu} x^{\mu})+c.c. ,
\end{eqnarray}
where $c.c.$ stands for the complex conjugation,
$C_{\mu\nu}$, a constant matrix, is the amplitude which satisfies the transverse-traceless conditions $q^{\mu}C_{\mu\nu}=0$ and $\eta^{\mu\nu}C_{\mu\nu}=0$,
$q^{\mu}$ is the wave-vector with $\eta^{\mu\nu}q_{\mu}q_{\nu}=0$ .

For the vector field $\phi^{(1)\mu}$, the linearized equation of Eq.~\eqref{eqv} gives
\begin{eqnarray}\label{}
\phi^{(1)\mu,\nu}_{~~~~~~,\nu}-\phi^{(1)\nu}_{~~~~,\mu,\nu}-(\mu^{(0)})^2 \phi^{(1)\mu}=0 .
\end{eqnarray}
According to the discussions presented in Ref.~\cite{green-moffat-toth}, the mass of vector field is approximately $2.8\times 10^{-28}$eV.
Since it is of the same order of the experimental bound for the photon mass, we will ignore the mass of the vector field hereafter. 
By imposing the Lorenz gauge $\phi^{(1)\mu}_{~~~~,\mu}=0$, the resultant equation governing the vector perturbation reads
\begin{eqnarray}\label{vector}
\phi^{(1)\mu,\nu}_{~~~~~~,\nu}=0.
\end{eqnarray}
The solution of the above equation takes the form
\begin{eqnarray}\label{solutiontovector}
\phi^{(1)\mu}=A^{\mu}\exp({i p_{\nu}x^{\nu}}) +c.c., 
\end{eqnarray}
where $p_{\nu}$ is the wave number satisfying $\eta^{\mu\nu}p_{\mu}p_{\nu}=-(\mu^{(0)})^2\approx 0$.

For the perturbations of scalar fields $G^{(1)}$ and $\mu^{(1)}$, the equations of motion are given by
\begin{eqnarray}\label{scalars}
{\psi^{,\mu}}_{,\mu}&=0,\\
\mu_{~~~~,\nu}^{(1),\nu}&=0.
\end{eqnarray}
The plane wave solution to Eq.~\eqref{scalars} are
\begin{eqnarray}\label{solutiontoscalars}
\psi=A_{\psi} \exp(i k_{\mu} x^{\mu})+c.c. ,\\
\mu^{(1)}=A_{\mu} \exp(i \tilde{k}_{\nu} x^{\nu})+c.c. ,
\end{eqnarray}
where $A_{\psi}$ and $A_{\mu}$ are the amplitudes,  while $k_{\mu}$ and $\tilde{k}_{\nu}$ represent the wave-vcectors. 
From Eqs.~\eqref{gw}, \eqref{vector}, and \eqref{scalars}, one also concludes that the transverse-traceless part $\overline{h}^{\mu\nu}$, vector perturbation $\phi^{(1)\mu}$, and scalar perturbation $\psi$ propogate at the speed of light~\cite{green-moffat-toth}.

\section{IV. The relative motion of neighboring particles}\label{relative-motion}

In this section, we set out to determine the relative motion of neighboring particles in Fermi normal coordinates.
These discussions regarding the dynamics of a test particle give rise to the polarization of the GWs in STVG.

Let us start by writing down the action of a point-like particle~\cite{moffat2006,moffat-toth} in STVG,
\begin{equation}
S=\int (-m-\omega q_5 \phi_{\mu}u^{\mu})d\tau
\end{equation}
where the integral is along the worldline of the test particle.
Here $\omega$ is a dimensionless coupling constant, $q_5$ represents the ``fifth force'' charge of the test particle.
It is related to the inertial mass of the particle, namely, $q_5=\kappa m$, where $\kappa$ is a constant. 
$u^{\mu}$ is the $4-$velocity of the test particle.
By varying the action, one obtains the equation of motion of a test particle where a force is exerted on the right-hand side of the equation~\cite{moffat2006,Mahmood} 
\begin{equation}\label{test}
a^{\mu}=\frac{d^2 x^{\mu}}{d\tau^2}+\Gamma^{\mu}_{~\alpha\beta}\frac{dx^{\alpha}}{d\tau}\frac{dx^{\beta}}{d\tau}=\omega \kappa B^{\mu}_{~\alpha}\frac{dx^{\alpha}}{d\tau},
\end{equation}
where $a^{\mu}$ is interpreted as the four-acceleration of the test particle.
By making use of the antisymmetric property of $B_{\mu\nu}$, it is easy to see that $a^{\mu}$ satisfies $u_{\mu}a^{\mu}=0$. 
However, since the four-acceleration $a^{\mu}$ is finite, the test particle does not free-fall along a geodesic.

By Eq.~\eqref{test}, one could investigate the relative displacement of the neighboring particles.
Consider an observer whose motion is described by the worldline $\sigma_{0}(\tau)$, where $\tau$ is the proper length of the worldline.
Let us use $u^{\mu}$ and $a^{\mu}$  to indicate his four-velocity and four-acceleration, respectively.
Moreover, let us introduce a spatial displacement of a neighboring particle with respect to the observer, denoted by $S^{\alpha}$.
The resultant equation for the relative four-acceleration of the test particles to the observer, $a^{\mu}_{rel}$, is found to be~\cite{Swaminarayan-Safko83,hou-gong}
\begin{equation}\label{relative-ab}
a^{\mu}_{rel}=-R_{\nu\alpha\beta}^{~~~~\mu}u^{\nu}S^{\alpha}u^{\beta}+S^{\alpha}a^{\mu}_{~;\alpha} ,
\end{equation}
where $u^{\mu}=(\frac{\partial}{\partial \tau})^{\mu}$ is the tangent vector of the trajectory of the test particle, chosen to be identical to that of the observer.

To study the relative motion of neighboring particles in the proper frame of reference of the observer~\cite{misner-thorne-wheeler}, one resorts to the Fermi normal coordinates of the latter regarding the worldline $\sigma_{0}(\tau)$.
Let us assume that the observer $\sigma_{0}(\tau)$ carries an orthonormal tetrad $\{ (e_{\hat{0}})^{a}=u^a,(e_{\hat{1}})^a, (e_{\hat{2}})^a,(e_{\hat{3}})^a$\}.
The latter satisfies the orthonormality condition $g_{ab}(e_{\hat{\mu}})^a(e_{\hat{\nu}})^a=\eta_{\hat{\mu}\hat{\nu}}$.
While the tetrad is Fermi-Walker transported along the worldline of the observer $\sigma_{0}(\tau)$, the line element near the observer's worldline reads~\cite{misner-thorne-wheeler}
\begin{eqnarray}\label{fermi-noraml}
ds^2=-(1+2a_{\hat{j}}x^{\hat{j}})d\tau^2+\delta_{\hat{j}\hat{k}}dx^{\hat{j}}dx^{\hat{k}}+O(|x^{\hat{j}}|^2),
\end{eqnarray}
where the indexes $\hat{j}, \hat{k}=1,2,3$. 
According to Eq.~\eqref{relative-ab}, the relative acceleration follows
\begin{eqnarray}\label{relative-accaleration}
a_{rel}^{\hat{j}}=-R_{\hat{0}\hat{k}\hat{0}}^{~~~~\hat{j}}S^{\hat{k}}+S^{\hat{k}}a^{\hat{j}}_{~;\hat{k}}
\end{eqnarray}
In the proper reference frame of the observer $\sigma_{0}(\tau)$, the non-vanishing components of the Christoffel connection are
\begin{eqnarray}\label{}
\Gamma^{\hat{0}}_{~\hat{0}\hat{j}}=\Gamma^{\hat{0}}_{~\hat{j}\hat{0}}=\Gamma^{\hat{j}}_{~\hat{0}\hat{0}}=a_{\hat{j}}.
\end{eqnarray}

Therefore, the relative acceleration can be expanded as
\begin{eqnarray}\label{relative-accaleration-expression}
a_{rel}^{\hat{j}}=u^{\hat{\mu}}\nabla_{\hat{\nu}}(u^{\hat{\nu}}\nabla_{\hat{\nu}}S^{\hat{j}})=\frac{d^2S^{\hat{j}}}{d\tau^2}+a^{\hat{j}}a_{\hat{k}}S^{\hat{k}} .
\end{eqnarray}
By further combining Eq.~\eqref{relative-accaleration} with Eq.~\eqref{relative-accaleration-expression}, one finds~\cite{hawking-ellis}
\begin{eqnarray}\label{deviation1}
\frac{d^2S^{\hat{j}}}{d\tau^2}=-R_{\hat{0}\hat{k}\hat{0}}^{~~~~\hat{j}}S^{\hat{k}}+S^{\hat{k}}a^{\hat{j}}_{;\hat{k}}-a^{\hat{j}}a_{\hat{k}}S^{\hat{k}}.
\end{eqnarray}

Up to the linear order in perturbation, Eq.~\eqref{deviation1} can be simplified to 
\begin{eqnarray}\label{deviation2}
\frac{d^2S^{\hat{j}}}{dt^2}=-R_{\hat{0}\hat{k}\hat{0}}^{~~~~\hat{j}}S^{\hat{k}}+S^{\hat{k}}\partial_{\hat{k}}a^{\hat{j}}=T_{\hat{k}}^{~\hat{j}} S^{\hat{k}} ,
\end{eqnarray}
where the matrix $T_{\hat{k}}^{~\hat{j}}=-R_{\hat{0}\hat{k}\hat{0}}^{~~~~\hat{j}}+\partial_{\hat{k}}a^{\hat{j}}$ . 
Following the discussions of Ref.~\cite{hou-gong}, to linear order the components of $T_{\hat{k}}^{~\hat{j}}$ in Fermi normal coordinates are equal to their counterparts in the TT coordinates, namely,
\begin{eqnarray}\label{T-matrix}
T_{\hat{k}}^{~\hat{j}}\approx T_{k}^{~j}=-R_{0k0}^{~~~~j}+a^{j}_{,k} ,
\end{eqnarray}
where $a^{j}=\omega \kappa B^{~j}_{0}$.

For a plane wave propagating along the $z$-axis, associated with the solution of the weak field in TT coordinates Eqs.~\eqref{solutiontogw} and \eqref{solutiontoscalars}, one has $h_{\mu\nu}^{(1)}=\overline{h}_{\mu\nu}(t-z)-\psi(t-z)\eta_{\mu\nu}$. 
As expected, $\overline{h}_{\mu\nu}$ induces the ``+'' and ``$\times$'' polarizations, and the massless scalar field is related to the breathing mode~\cite{hou-gong-liu,hou-gong18}.

Now let us investigate the relative motion of neighboring particles caused by the vector field $\phi^{\mu}$.
For simplicity, we take $\overline{h}_{\mu\nu}=0$ and $\psi=0$.
Concerning the solution of $\phi^{(1)\mu}$ in TT gauge Eq.~\eqref{solutiontovector},  we write it down as $\phi^{(1)\mu}=A_{\phi}^{\mu}\cos(k_{\phi}t-k_{\phi}z)$.
Here, the amplitudes $A_{\phi}^{\mu}=\{A^3, A^1, A^2, A^3\}$ are defined in Lorenz gauge.
Also, the matrix $a^{j}_{,k}$ reads
\begin{eqnarray}\label{}
\left[\begin{matrix} 0&0&-A^{1} k_{\phi}^2\cos(k_{\phi}t-k_{\phi}z)\nonumber\\ 
0&0&-A^{2} k_{\phi}^2\cos(k_{\phi}t-k_{\phi}z)\nonumber\\ 0&0&0\\
\end{matrix}\right].
\end{eqnarray}

Putting all pieces together, it is straightforward to show that the system of equations for the relative motion can be simplified to
\begin{subequations}\label{geodesicdeviation}
\begin{align}
&\ddot{\delta x}+\omega \kappa A^{1}k_{\phi}^2\cos(k_{\phi}t-k_{\phi}z)\delta z=0,\label{gd1}\\
&\ddot{\delta y}+\omega \kappa A^{2}k_{\phi}^2\cos(k_{\phi}t-k_{\phi}z)\delta z=0,\label{gd2}\\
&\ddot{\delta z}=0,\label{gd3}
\end{align}
\end{subequations}
where the dotted ``.'' means the derivative with respect to $t$, and $A^{1}$ and $A^{2}$ are the components of the amplitude of  plane wave solution $\phi^{(1)}$.  
Due to Eq.~\eqref{gd3}, if the $z$-components of the initial relative velocity and acceleration are zero, then the deviation $\delta z$ will remain unchanged, thus labeled by $\delta z_0$. 
The solutions to Eq.~\eqref{geodesicdeviation} are 
\begin{eqnarray}\label{geodesic-deviation-solutions}
\delta x&&=\delta x_0+\omega \kappa A^1 k_{\phi}^2\delta  z_0 [\cos(k_{\phi}t-k_{\phi}z)-\cos(k_{\phi}t_0-k_{\phi}z)] ,\nonumber\\
\delta y&&=\delta y_0+\omega \kappa A^2 k_{\phi}^2 \delta z_0 [\cos(k_{\phi}t-k_{\phi}z)-\cos(k_{\phi}t_0-k_{\phi}z)] ,\nonumber\\
\delta z&&=\delta z_0 ,\nonumber
\end{eqnarray}
where $\delta x_0(x,y,z)$, $\delta y_0(x,y,z)$, and $\delta z_0(x,y,z)$ are the components of the initial relative displacement.
The above equation shows that there exist two transverse oscillations induced by the vector field.

To conclude, there are a total of five polarization modes.
Two transverse modes and one breathing mode are induced by the metric tensor, namely, the TT part and the trace part excited by the scalar field. 
Another two transverse modes are associated with the vector field. 
The results on the polarization in this section are in agreement with those discussed in Ref.~\cite{green-moffat-toth}.

\section{V. The Stress-energy Pseudo-tensor}\label{pseudo-tensor}

In this section, we evaluate the GW SET by using the perturbed field equation method, which was first developed in general relativity by Isaacson~\cite{isaacson1,isaacson2}. 
We first expand the fields to the second order as follows
\begin{subequations}\label{fieldsperts}
\begin{align}
&g_{\mu\nu}=\eta_{\mu\nu}+h^{(1)}_{\mu\nu}+h^{(2)}_{\mu\nu},\label{fieldsperts1}\\
&\phi^{\mu}=\phi^{(1)\mu}+\phi^{(2)\mu},\label{fieldsperts2}\\
&G=G^{(0)}+G^{(1)}+G^{(2)},\label{fieldsperts3}\\
&\mu= \mu^{(0)}+\mu^{(1)}+\mu^{(2)},\label{fieldsperts4}
\end{align}
\end{subequations}
where $\eta_{\mu\nu}$ is the Minkowski metric.
The superscript ``$(n)=(1),(2)$'' denotes the order of the perturbation.
For instance, $h^{(1)}_{\mu\nu}$ and $h^{(2)}_{\mu\nu}$ are the first and second-order perturbation of the metric, respectively. 
By substituting Eq.~\eqref{fieldsperts}, one may systematically derive the linearized equation order by order. 
In fact, the equations of the first order are precisely Eqs.~\eqref{gw}, \eqref{vector}, and \eqref{scalars}, as we have obtained before.

Now, by expanding Eq.~\eqref{fieldsequations} to the second-order, one obtains the expressions for the effective SET $T^{GW,eff}_{\mu\nu}$ as follows.
%We delegate the details of the derivation to the Appendix.
\begin{eqnarray}\label{TGWeff}
G_{\alpha\beta}(h^{(2)})&=&8\pi G^{(0)}(T^{GW,eff}_{\alpha\beta}+t_{\alpha\beta}^{\mu}),\nonumber\\
8G_{\mu\nu}(h^{(2)})&=&-\left<X_{\mu\nu}\left[(h^{(1)})^2, (\phi^{(1)})^2,(G^{(1)})^2,h^{(1)}G^{(1)},G^{(2)}\right]\right> ,
\end{eqnarray}
where $t_{\alpha\beta}^{\mu}$ denotes the energy-momentum tensor of the vector field perturbation $\mu^{(1)}$, the angled-brackets stand for short-wavelength averaging~\cite{isaacson1,isaacson2}. 
As discussed in the previous section, the GWs in STVG theory are propagated regarding the transverse-traceless tensor $h^{TT}_{\mu\nu}$ of the metric, $G^{(1)}$ of the scalar, and $\phi^{(1)\mu}$ of the vector field perturbations.

Since the resultant SET is averaged, one may utilize the integration by parts to eliminate the boundary terms. 
Furthermore, by plugging in the equation of motion of the first order perturbations, Eqs.~\eqref{gw}, \eqref{vector}, and \eqref{scalars}, while imposing the transverse-traceless gauge for $\overline{h}_{\mu\nu}$, one finds
\begin{eqnarray}\label{septpert}
&&32\pi G^{(0)}T^{GW,eff}_{\mu\nu}=\left<\overline{h}_{\gamma\delta,\mu}^{TT}\overline{h}^{\gamma\delta}_{TT,\nu}+(32\pi-6)\psi_{,\mu}\psi_{,\nu}+32\pi G^{(0)}\phi^{(1)\gamma}_{~~~~,\mu}\phi^{(1)}_{\gamma,\nu}\right>
\end{eqnarray}
In particular, if the vector ($\phi^{\mu}$) and scalar ($\mu$) fields vanish, it is readily to show that the action Eq.~\eqref{actionterms} falls back to the Brans-Dicke theory by redefining $\tilde{\phi}=1/G $. 
Also, the result given by Eq.~\eqref{septpert} is consistent with the GW SET of Brans-Dicke theory in~\cite{will1993,saffer-yunes-yagi}.

With the plane wave solutions Eqs.\eqref{solutiontogw}, \eqref{solutiontovector}, and \eqref{solutiontoscalars} at our disposal, we proceed to calculate the stress-energy tensor for a single plane wave. 
The resultant plane waves read
\begin{eqnarray}\label{}
\overline{h}_{\alpha\beta}^{TT}&=&C_{\alpha\beta}\cos(q_{\lambda}x^{\lambda}),\\
\phi^{(1)\alpha}&=&A_{\phi}^{\alpha}\cos(p_{\lambda}x^{\lambda}),\\
\psi&=&A_{\psi}\cos(k_{\lambda}x^{\lambda}).
\end{eqnarray}
By imposing the $\sin^2$ term over several wavelengths is equal to $1/2$, the stress-energy tensor \eqref{septpert} is then
\begin{eqnarray}\label{}
T^{GW,eff}_{\alpha\beta}=\frac{1}{64\pi G^{(0)}}\left[q_{\alpha}q_{\beta}C^{\mu\nu}C_{\mu\nu}+(32\pi-6)k_{\alpha}k_{\beta}A_{\psi}^2+32\pi(G^{(0)})^2p_{\alpha}p_{\beta}A^{\phi}_{\mu}A_{\phi}^{\mu} \right]
\end{eqnarray}
For the plane wave propagating along the $z$ direction, so that 
\begin{eqnarray}\label{}
p_{\lambda}=(-p,0,0,p),~~q_{\lambda}=(-q,0,0,q),~~k_{\lambda}=(-k,0,0,k).
\end{eqnarray}
In the TT gauge, the only nonvanishing components of the matrix $C_{\mu\nu}$ are $C_{11}=-C_{22}=h_{+},~C_{12}=C_{21}=h_{\times}$. 
Subsequently, one obtains
\begin{eqnarray}\label{gwsetlast}
&&T^{GW,eff}_{\alpha\beta}=\frac{\pi}{8 G^{(0)}}\left[f_1^2\left(h_{+}^2+h_{\times}^2\right)\right.\nonumber\\ &&\left.+(16\pi-3)(f_3)^2A_{\psi}^2+16\pi G^{(0)}(f_2)^2\left((A^{1})^2+(A^{2})^2\right) \right]e_{\alpha\beta},
\end{eqnarray}
where $f_1=\frac{q}{2\pi},f_2=\frac{p}{2\pi},f_3=\frac{k}{2\pi}$ are the ordinary frequencies, and 
\begin{eqnarray}\label{}
e_{\alpha\beta}=\left[\begin{matrix} 1&0&0&-1\nonumber\\ 
0&0&0&0\\ 0&0&0&0\\-1&0&0&1
\end{matrix}\right].
\end{eqnarray}

Eq.\eqref{gwsetlast} is obtained in the TT coordinate. 
Similar to the discussions in Ref.~\cite{hou-gong}, the Fermi normal coordinates differ from the TT coordinates for quantities of order one.
Since the field perturbations $\overline{h}_{\alpha\beta}^{TT}$, $\phi^{(1)\mu}$, and $\psi$ are all of the linear order, any change in their components due to the coordinate transformation is of the second order in perturbations.
Therefore, the SET in the proper reference frame of the observer $\sigma_0(\tau)$ remains the same as Eq.~\eqref{gwsetlast}. 
As the vector and scalar fields vanish, Eq.~\eqref{gwsetlast} falls back to its counterpart in General Relativity.

\section{VI. Extension to sources with non-negligible self-gravity}\label{non-negligible self gravity}

For systems with weak gravity whose dynamics are dominated by self-gravity, the above procedure of obtaining the linearized gravitational field equation is no longer applicable.
This was first pointed out by Eddington.
Examples of such systems are binary star systems or those with non-linear GW memory effect~\cite{payne,blanchet-damour,christodoulou,thorne}.
In this case, it is still possible to extend the derivation to encompass systems with non-negligible self-gravity.
In this section, following Ref.~\cite{thorne1975}, we derive the exact, nonlinear gravitational field equations Eqs.~\eqref{fieldsequations} in an arbitrary coordinate system regarding the Landau-Lifshitz formalism~\cite{misner-thorne-wheeler,landau-lifshitz}. 
Also, the GW SET is evaluated for a second time in this part by employing the above formalism.

One first defines 
\begin{eqnarray}\label{mathg}
\mathbf{g}^{\mu\nu}&&=(-g)^{1/2}g^{\mu\nu}, (-g)=-{\mathrm {det}}\|g_{\mu\nu}\|=-{\mathrm {det}}\|\mathbf{g}^{\mu\nu}\|,\\
\mathcal{H}^{\mu \gamma\nu \delta}&&=G^{-2}(\mathbf{g}^{\mu\nu}\mathbf{g}^{\gamma \delta}-\mathbf{g}^{\gamma \nu}\mathbf{g}^{\mu\delta}),
\end{eqnarray}
By making use of the above notations, Eq.~\eqref{fieldsequations} can be rewritten as 
\begin{eqnarray}\label{H4}
\mathcal{H}_{~~~~~,\gamma\delta}^{\alpha \gamma\beta \delta}=-\frac{16\pi g}{G}(T_M^{\alpha\beta}+t^{\alpha\beta}+t_{\mu}^{\alpha\beta}) .
\end{eqnarray}

The symmetric properties of $\mathcal{H}^{\mu \gamma\nu \delta}$ ensure that the left-hand side of Eq.~\eqref{H4} vanishes upon differentiation with respect to $\mu$ or $\nu$.
To be specific, we have 
\begin{eqnarray}
\left[(-g)(T_M^{\alpha\beta}+t^{\alpha\beta}+t^{\alpha\beta}_{\mu})/G\right]_{,\alpha}=0 .
\end{eqnarray}
By a rather lengthy calculation, one arrives at the following relations for $t^{\mu\nu}$
\begin{eqnarray}\label{tt-relaxed}
t^{\mu\nu}&=&t_{LL}^{\mu\nu}+t_{f}^{\mu\nu},\\
t^{\alpha\beta}_{\mu}&=&\frac{1}{2G\mu^2}(g^{\alpha\beta }\mu_{;\gamma}\mu^{;\gamma}-2\mu^{;\alpha}\mu^{;\beta}),\\
t_f^{\mu\nu}&=&\frac{1-4\pi }{4\pi G^3}G^{;\mu}G^{;\nu}+\frac{g^{\mu\nu}G^{;\gamma}_{~~;\gamma}-G^{;\mu\nu}}{8\pi G^2}+g^{\mu\nu}(2\pi-1)\frac{G_{;\gamma}G^{;\gamma}}{4\pi G^3}\nonumber\\
&+&\frac{1}{2}g^{\mu\nu}\phi^{\gamma;\delta}B_{\gamma\delta}+B^{\mu\gamma}\phi_{\gamma}^{~;\nu}+\phi^{\nu}B^{\gamma\mu}_{~~;\gamma}+\phi^{\mu}B^{\gamma\nu}_{~~;\gamma}\nonumber\\
&+&\phi^{\nu;\gamma}B_{\gamma}^{~\mu}+(g^{\mu\nu}g^{\gamma\delta}-g^{\mu\gamma}g^{\nu\delta})(3G_{,\gamma}G_{,\delta}-G_{,\delta\gamma}G)/8\pi G^3\nonumber\\
&+&g_{\epsilon\zeta(,\delta}G_{,\gamma)}(g^{\mu\nu}g^{\gamma[\epsilon }g^{\delta]\zeta}+g^{\mu\gamma}g^{\nu[\delta}g^{\zeta]\epsilon}+g^{\mu\epsilon}g^{\delta[\nu}g^{\gamma]\zeta})/2\pi G^2,
\end{eqnarray}
where $B^{\mu\nu}=\partial^{\mu}\phi^{\nu}-\partial^{\nu}\phi^{\mu}=\phi^{\nu;\mu}-\phi^{\mu;\nu}$, $t^{\mu\nu}_{LL}$ is known as the Landau-Lifshitz pseudotensor~\cite{landau-lifshitz}, defined by 
\begin{eqnarray}
16\pi G\sqrt{-g}t^{\alpha\beta}_{LL}&=&\mathbf{g}^{\alpha\beta}_{~~,\gamma}\mathbf{g}^{\gamma\delta}_{~~,\delta}-\mathbf{g}^{\alpha\gamma}_{~~,\gamma}\mathbf{g}^{\beta\delta}_{~~,\delta}+\frac{1}{2}g^{\alpha\beta}g_{\gamma\delta}\mathbf{g}^{\gamma\kappa}_{~~,\epsilon}\mathbf{g}^{\delta\epsilon}_{~~,\kappa}\nonumber\\&-&2g_{\delta\epsilon}g^{\gamma(\alpha}\mathbf{g}^{\beta)\epsilon}_{~~~,\kappa}\mathbf{g}^{\delta\kappa}_{~~,\gamma}\nonumber+g_{\gamma\delta}g^{\epsilon\kappa}\mathbf{g}^{\alpha\gamma}_{~~,\epsilon}\mathbf{g}^{\beta\delta}_{~~,\kappa}\nonumber\\&+&\frac{1}{8}(2g^{\alpha\gamma}g^{\beta\delta}-g^{\alpha\beta}-g^{\alpha\beta}g^{\gamma\delta})(2g_{\epsilon\kappa}-g_{\kappa\lambda}g_{\epsilon\delta})\mathbf{g}^{\epsilon\delta}_{~~,\gamma}\mathbf{g}^{\kappa\lambda}_{~~,\delta}.
\end{eqnarray}

For the case of weak field, we define the potentials
\begin{eqnarray}\label{tildeh}
G&=&G^{(0)}+\delta G,\\
\mathbf{g}^{\mu\nu}&=&(\eta^{\mu\nu}-\tilde{h}^{\mu\nu})G/G^{(0)},\\
g^{\mu\nu}&=&\eta^{\mu\nu}-h^{\mu\nu},
\end{eqnarray}

By combining Eq.~\eqref{mathg} and Eq.~\eqref{tildeh}, we obtain
\begin{eqnarray}
\tilde{h}^{\mu\nu}&=&G\sqrt{-g}(\eta^{\mu\nu}-h^{\mu\nu}).\label{tildeh}
\end{eqnarray}

It is convenient to work in a particular coordinate system introduced by the de Donder, namely, the harmonic gauge condition $\tilde{h}^{\mu\nu}_{~,\mu}=0$, we have
\begin{eqnarray}
\mathcal{H}_{~~~~~,\gamma\delta}^{\mu \gamma\nu \delta}=-\left(\square_{F}\tilde{h}^{\mu\nu}+\tilde{h}^{\mu\gamma}_{~,\delta}\tilde{h}^{\nu\delta}_{~,\gamma}-\tilde{h}^{\gamma\delta}\tilde{h}^{\mu\nu}_{~,\gamma\delta}\right)/(G^{(0)})^{2},
\end{eqnarray}
where $\square_{F}\equiv \eta^{\mu\nu}\partial_{\mu}\partial_{\nu}$ is the flat-spacetime wave operator. 

By putting all the pieces together, the resultant relaxed gravitational equations takes the form
\begin{eqnarray}\label{rewrittenGR}
\square_{F} \tilde{h}^{\mu\nu}=-16\pi (G^{(0)})^{2} \tau^{\mu\nu} .
\end{eqnarray}
The source on the right-hand side is interpreted as the ``effective'' SET, namely, 
\begin{eqnarray}\label{rewrittenGRtensor}
\tau^{\mu\nu}=(-g)(G^{(0)})^{2}(T_M^{\mu\nu}+t^{\mu\nu}_{f}+t^{\mu\nu}_{\mu})/G+ (16\pi)^{-1}(\tilde{h}^{\mu\gamma}_{~~,\delta}\tilde{h}^{\nu\delta}_{~~,\gamma}-\tilde{h}^{\gamma\delta}\tilde{h}^{\mu\nu}_{~~,\gamma\delta}).
\end{eqnarray}
If, in the action Eq.~\eqref{action}, the vector field $\phi^{\mu}$ and scalar field $\mu$ vanish, by redefining $\tilde{\phi}=1/G $, one observes that the STVG theory falls back to Brans-Dicke theory, while Eq.~\eqref{rewrittenGR} also reduces to its counterpart in Brans-Dicke theory~\cite{du-nishizawa}.
For the latter, the authors studied the nonlinear memory effect and discovered two new types of memory. 

From Eq.~\eqref{tt-relaxed}, if one considers the field perturbations as those given in Eq.~\eqref{fieldsperts}, the SET of GWs can be obtained.
This can be achieved by first expanding Eq.~\eqref{tt-relaxed} to the second order.
Eq.~\eqref{tildeh} can be further expanded to read
\begin{eqnarray}
\tilde{h}^{\mu\nu}&=&h^{\mu\nu}-\frac{1}{2}\eta^{\mu\nu}+\frac{\delta G}{G^{(0)}}\eta^{\mu\nu}+O[(h)^2,(\delta G)^2,h\delta G] .
\end{eqnarray}
We note that, to the first order, the above equation is precisely Eq.~\eqref{newpert}.

Subsequently, one integrates relevant terms by parts to eliminate the boundary term, plugs in the equation of motion of the first order perturbations, and imposes the transverse-traceless gauge for $\tilde{h}_{\mu\nu}$.
Eventually one finds
\begin{eqnarray}\label{septll}
32\pi G^{(0)}T^{GW,eff}_{\mu\nu}=\left<\tilde{h}_{\gamma\delta,\mu}^{TT}\tilde{h}^{\gamma\delta}_{TT,\nu}+(32\pi-6)\psi_{,\mu}\psi_{,\nu}+32\pi G^{(0)}\phi^{(1)\gamma}_{~~~~,\mu}\phi^{(1)}_{\gamma,\nu}\right> .
\end{eqnarray}
We note that Eq.~\eqref{septll} is identical to the form of the GW SET found previously in Eq.~\eqref{septpert}.

\section{VII. Concluding remarks}\label{conclusion}

To summarize, the present paper is dedicated to studying various aspects of the GWs in a modified theory of gravity, namely, the STVG theory.
The latter is an alternative gravity theory characterized by the exchange of dynamical scalar fields.
As the model is shown to be in good agreement in the context of the weak field, it is meaningful to further investigate its validity for the strong field, regarding GWs.  
In particular, we analyze the polarization in terms of geodesic deviation equation.
The SET is explored by using both the perturbed equation method and the Landau-Lifshitz formalism.
The obtained GW SET from both methods is shown to be identical, although the latter is understood to be appropriate to the systems where self-gravitation is essential.
Also, the relaxed field equation is obtained by employing the latter method.
By studying the relative motion of the test particles, we derive the polarization modes of the GWs. 

Although not addressed in the present study, it is meaningful to eventually compare the properties of GWs, namely, the polarization and energy propagation, against experimental observations.
Further study in this direction is in progress.

\acknowledgments
We thank De-Chang Dai for the valuable discussions. 
This research is supported in part by the Major Program of the National Natural Science Foundation of China under Grant No. 11690021 and the National Natural Science Foundation of China under Grant No. 11475065 and 11505066. 
We also gratefully acknowledge the financial support from
Funda\c{c}\~ao de Amparo \`a Pesquisa do Estado de S\~ao Paulo (FAPESP),
Funda\c{c}\~ao de Amparo \`a Pesquisa do Estado do Rio de Janeiro (FAPERJ),
Conselho Nacional de Desenvolvimento Cient\'{\i}fico e Tecnol\'ogico (CNPq),
and Coordena\c{c}\~ao de Aperfei\c{c}oamento de Pessoal de N\'ivel Superior (CAPES).
The results concerning the GW SET and relaxed field equation are obtained by using the tensor-algebra bundle xAct~\cite{xAct}.

\end{document}